\documentclass[sigconf,screen]{acmart}

\usepackage{booktabs}
\usepackage{multirow}
\usepackage{makecell}
\usepackage{mdframed}
\usepackage{csquotes}
\usepackage{enumitem}
\usepackage{colortbl}

\definecolor{ColorPink}{HTML}{FF8DAE}
\definecolor{ColorOrange}{HTML}{EE8866}
\definecolor{ColorBlue}{HTML}{77AADD}
\definecolor{ColorStronglyDisagree}{HTML}{DD3D2D}
\definecolor{ColorSomewhatDisagree}{HTML}{FF8DAE}
\definecolor{ColorNeutral}{HTML}{DDDDDD}
\definecolor{ColorSomewhatAgree}{HTML}{99DDFF}
\definecolor{ColorStronglyAgree}{HTML}{77AADD}
\definecolor{ColorGrayBackground}{HTML}{EEEEEE}

\definecolor{sessions}{RGB}{238,119,51}
\definecolor{deadlines}{HTML}{0d88e6}
\newcommand\Finding[1]{
	\noindent 
	\begin{mdframed}[
		linewidth=1pt,
		leftmargin=0pt,
		backgroundcolor=gray!10,
		linecolor=gray!20]
		{#1}
	\end{mdframed}
}

\newcommand{\TablePlot}[1]{\includegraphics[trim={0 0 0 0}, clip, width=4cm, height=0.36cm, keepaspectratio=false]{#1}}

\newcommand{\rotated}[1]{\begin{tabular}{@{}c@{}}\rotatebox[origin=c]{90}{#1}\end{tabular}}

\newcommand{\specialmidruleTop}{\arrayrulecolor[HTML]{77AADD}
\specialrule{0.5pt}{4pt}{0pt}\arrayrulecolor{black}}
\newcommand{\specialmidruleBottom}{\arrayrulecolor[HTML]{77AADD}
\specialrule{0.5pt}{0pt}{2pt}\arrayrulecolor{black}}

\copyrightyear{2026}
\acmYear{2026}
\setcopyright{cc}
\setcctype{by}
\acmConference[FSE Companion '26]{34th ACM Joint European Software Engineering Conference and Symposium on the Foundations of Software Engineering}{July 05--09, 2026}{Montreal, QC, Canada}
\acmBooktitle{34th ACM Joint European Software Engineering Conference and Symposium on the Foundations of Software Engineering (FSE Companion '26), July 05--09, 2026, Montreal, QC, Canada}
\acmDOI{10.1145/3803437.3805785}
\acmISBN{979-8-4007-2636-1/2026/07}

\begin{document}

\title[Harnessing Hype to Teach Empirical Thinking]{Harnessing Hype to Teach Empirical Thinking: An~Experience~With~AI~Coding~Assistants}

\author{Marvin Wyrich}
\authornote{Both authors contributed equally to this work.}
\email{wyrich@cs.uni-saarland.de}
\orcid{0000-0001-8506-3294}
\affiliation{%
  \institution{Saarland University}
  \city{Saarbr{\"u}cken}
  \country{Germany} 
}

\author{Norman Peitek}
\authornotemark[1]
\email{peitek@cs.uni-saarland.de}
\orcid{0000-0001-7828-4558}
\affiliation{%
  \institution{Saarland University}
  \city{Saarbr{\"u}cken}
  \country{Germany} 
}

\author{Kallistos Weis}
\orcid{0009-0001-9120-1670}
\email{kallistos@cs.uni-saarland.de}
\affiliation{%
  \institution{Saarland University}
  \city{Saarbr{\"u}cken}
  \country{Germany} 
}

\author{Sven Apel}
\orcid{0000-0003-3687-2233}
\email{apel@cs.uni-saarland.de}
\affiliation{%
  \institution{Saarland University}
  \city{Saarbr{\"u}cken}
  \country{Germany} 
}

\renewcommand{\shortauthors}{Wyrich and Peitek et al.}

\begin{abstract}
Software engineering students often struggle to appreciate empirical methods and hypothesis-driven inquiry, especially when taught in theoretical terms. This experience report explores whether grounding empirical learning in hype-driven technologies can make these concepts more accessible and engaging.    
We conducted a one-semester seminar framed around the currently popular topic of AI coding assistants, which attracted unusually high student interest. The course combined hands-on sessions using AI coding assistants with small, student-designed empirical studies.
    
Classroom observations and survey responses suggest that the hype topic sparked curiosity and critical thinking. Students engaged with the AI coding assistants while questioning their limitations---developing the kind of empirical thinking needed to assess claims about emerging technologies.
Key lessons: (1) Hype-driven topics can lower barriers to abstract concepts like empirical research; (2) authentic hands-on development tasks combined with ownership of inquiry foster critical engagement; and (3) a single seminar can effectively teach both technical and research skills.
\end{abstract}

\ccsdesc[500]{Social and professional topics~Computing education}
\keywords{Empirical Software Engineering, Inquiry-Based Learning, Critical Thinking, Technology Hype, Experience Report}

\maketitle

\section{Introduction}

\enquote{\textit{I just wanted to try Copilot, and now I'm doing a user study,}} one student laughs. 
We cannot help but smile---because in that moment, we feel a little caught.
Our primary goal was to teach empirical thinking: how to form hypotheses, design studies, and reason soundly from data.
That goal has not changed from previous seminars offered by our chair.
What changed was the framing.
Instead of asking students to empirically investigate well-established tools or practices, we invited them to explore an emerging technology they were already eager to try.
A bit of sugar on the broccoli, one might say.
The result? Record application numbers.

Understanding the appeal of this seminar starts with the hype surrounding AI coding assistants.
AI coding assistants---tools that use large language models (LLMs) to suggest code and complete programming tasks---have rapidly entered the software industry.
GitHub Copilot, one of the most prominent examples, integrates seamlessly into popular development environments and provides real-time code suggestions based on natural-language prompts or partially written code~\cite{GitHubCopilotWebsite}.
Recent advances in LLM research have fueled interest in these tools and their potential to support everyday development work.
But the excitement surrounding AI coding assistants is part of a familiar pattern: In software engineering, trending technologies have always had strong appeal~\cite{Broy:2025:Hype,Antinyan:2026:Fashion}.
Sometimes that appeal stems from genuine curiosity or the hope for increased productivity, sometimes from the belief that being familiar with popular tools signals employability.
The phenomenon of \emph{Résumé-Driven Development} illustrates this tendency: developers and employers alike are drawn to what is perceived as current and in demand, even when those choices are not always the most suitable for the task at hand~\cite{Fritzsch:2021:RDD, Fritzsch:2023:RDD}.

As with many industry trends, this one is quickly making its way into the classroom, where educators face the challenge of deciding whether and how to incorporate AI coding assistants into the curriculum~\cite{DagstuhlGenAI}.
In just the past three years, the topic has attracted so much academic attention that several secondary studies have already begun synthesizing the emerging body of research~\cite{Prather2023, pirzado2024navigating}.
Research in this space primarily explores the opportunities, challenges, and risks of integrating generative AI into higher education~\cite{Becker2023, Choudhuri2025, Kirova2024, Prather2024, Galej2025, Rodrigues2025}, discusses institutional strategies for responding to these developments~\cite{Lau2023, Berrezueta:2025:CodersCritics}, documents early classroom observations~\cite{kazemitabaar2024codeaid, Margulieux2024, Yilmaz2023}, and provides initial guidelines on how to include AI coding assistants in teaching~\cite{Mahon2024, Vierhauser:2024:AIinEdu}.
Some contributions even argue from an industry perspective that such tools should be actively taught in university settings~\cite{Bull2024}. 

While AI coding assistants are just one facet of the broader generative AI movement in education, they offer a particularly tangible entry point---one that invites technical exploration and also raises important questions about how such tools should be evaluated in practice.
For example, many claims surrounding tools such as GitHub Copilot center on increased productivity or improved developer experience~\cite{GitHubCopilotWebsite}.
But how can we know whether such claims hold up under scrutiny, and whether it is worth investing time and resources into these tools?
Teaching students how to answer such questions remains a fundamental but challenging task.
Empirical methods are central to software engineering research~\cite{Mendez:2025:HandbookTeaching} and practice~\cite{Wilson:2025:Workshop}, yet often perceived as abstract or dry, especially when taught in isolation from concrete problems.

We wondered: Can we harness the students' excitement around hype-driven technologies such as AI as a hook---while fostering critical, evidence-based thinking? 
In this experience report, we describe a seminar that set out to do exactly that.
Students explored AI coding assistants not just as users, but as investigators, learning how to design studies, formulate hypotheses, and interpret results along the way.
Through this seminar, we contribute empirical evidence for the pedagogical value of hype-driven topics in fostering critical reflection on scientific methodology. 
We also share three publicly available teaching artifacts: hands-on materials for learning to use GitHub Copilot, resources for evaluating students' experience and learning, and the raw data from our seminar evaluation.\footnote{\url{https://github.com/brains-on-code/coding-assistants-empirical-thinking}}

In the following, we outline the seminar's design and structure (Section~\ref{sec:design}), share impressions from our teachers' perspective to illustrate student engagement and learning during the course (Section~\ref{sec:impressions}), present the students' perspective from a post-seminar survey exploring how the hype topic shaped their experiences and perceptions (Section~\ref{sec:survey}), reflect on broader lessons and implications for teaching empirical methods in fast-moving technological landscapes (Section~\ref{sec:lessons}), and conclude with a short summary (Section~\ref{sec:conclusion}).

\section{Course Design}
\label{sec:design}

Seminars play a central role in both the bachelor’s and master’s computer science programs at our university.
They are semester-long courses, typically limited to 12--20 participants, with most chairs offering one or two seminars each term.
As part of their degree requirements, students must complete, at least, two seminars in the bachelor’s program and, at least, one in the master’s program.
While general learning objectives are defined at the departmental level, the detailed course design is left to the offering chair, resulting in substantial variation across seminars.

Students select from a department-wide list of available seminars, each announced with a title and description outlining the scope and expected deliverables.
Applications include a brief motivation letter, which is reviewed by the responsible chairs to assess fit.
A central allocation system then assigns students to seminars based on their preferences, evaluations, and global balancing criteria.

In what follows, we describe the specific design of our seminar on AI coding assistants, including its overall structure, learning objectives, participants, and the activities that shaped the semester.

\begin{figure*}[t]
    \centering
    \includegraphics[width=1\linewidth]{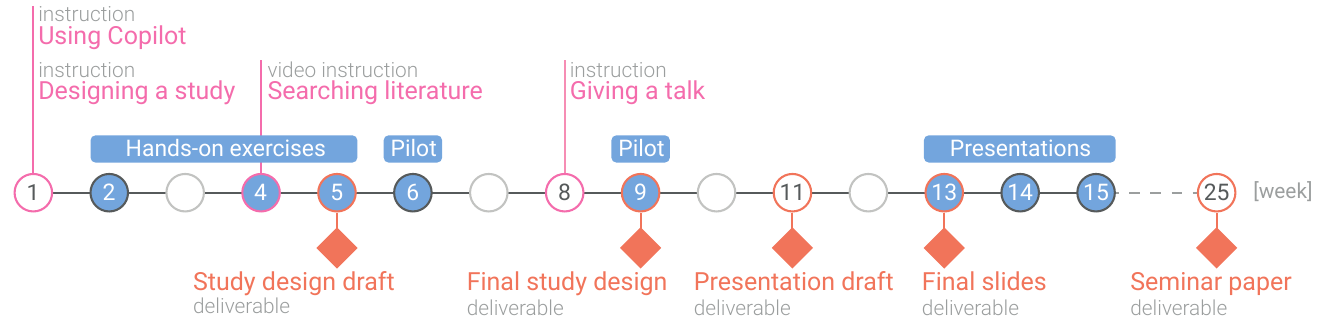}
    \caption{Schematic overview of the seminar structure, showing instructional sessions where content was taught to students (\textcolor{ColorPink}{pink}), sessions involving student activities (\textcolor{ColorBlue}{blue}), and milestones with deliverables due at each point (\textcolor{ColorOrange}{orange}). The timeline is organized by week; numbers indicate weeks with an on-site session or a milestone deliverable.}
    \Description{Schematic timeline of our seminar structured by week. Colored elements distinguish instructional sessions, student activity sessions, and milestone deliverables. Only selected weeks are labeled, indicating when on-site sessions or deadlines occur.}
    \label{fig:CourseStructure}
\end{figure*}

\subsection{Overall Structure}
We provide an overview of the seminar structure in~\autoref{fig:CourseStructure}.
The seminar comprised ten sessions of $90$~minutes each.
The first session introduced the overall structure of the seminar, explained all deliverables in detail, included a short lecture on designing empirical studies, and demonstrated how to use GitHub Copilot in Visual Studio Code.
Sessions $2$--$4$ were hands-on exercises in which students worked on programming tasks with and without GitHub Copilot. Each exercise emphasized a different use case---code generation, code explanation and refactoring, and debugging---allowing students to directly experience the tool's strengths and limitations in varied contexts.
At the same time, we designed these sessions to spark ideas for students' own research questions by encouraging them to critically reflect on their observations.
Sessions $5$ and $7$ focused on students' own planned studies: Each student tested their study design with a peer and reciprocally participated in a peer’s study.
Session $6$ covered principles of scientific presentations, followed by a Q\&A session focused on refining study designs after the first pilot.
In sessions $8$--$10$, students presented their research questions, study designs, initial findings and insights of their pilot study, as well as reflections on their empirical work.

After the in-class phase, students had $9.5$ weeks to write a seminar paper.
The paper outlines the motivation for their research question and study, reviews related work, and documents refinements based on the two pilot iterations, including preliminary results.
We encouraged students to critically engage with their own study design, for instance, by reflecting on validity aspects and documenting potential limitations alongside their findings. 
This element of \emph{active learning} has been shown to help in deepening students' understanding of empirical research~\cite{Meireles2024}.

\subsection{Learning Objectives and Principles}

We designed the seminar according to the principles of constructive alignment~\cite{BiggsTang2011TQFLAU}, ensuring that teaching activities, assessments, and learning objectives were mutually supportive.
By the end of the course, students were expected to be able to independently design, refine, and execute an empirical study, reflecting critically on their methodological choices and potential validity threats based on recommendations of teaching empirical studies~\cite{Molleri:2024:TeachingResearchDesign}.
They were also expected to demonstrate ownership of their research by exploring a current, relevant topic and making independent design choices rather than replicating established studies.
Finally, students were expected to communicate their findings clearly through structured scientific writing, integrating feedback from staged drafts.

We aligned our grading with these objectives and based it on the seminar paper, oral presentation, and active participation.
All seminar instructors jointly reviewed and discussed each part, reaching consensus to ensure consistent evaluation and alignment with the intended learning outcomes.

\subsection{Participants and Instructors}
At our department, seminars limit enrollment to maintain close supervision.
In our chair, we aim for a staff-to-student ratio of approximately $1{:}1$ to $1{:}2$.
Of $168$ applicants for our seminar, $18$ students were selected (by the departments' distribution system), of whom $13$ concluded the seminar.
We paired these $13$ students with nine instructors (doctoral and postdoctoral researchers) from our chair, each student receiving one-to-one feedback across key milestones, including guidance on study design and presentation.
We consider this close supervision highly beneficial for fostering student learning, though we recognize that not all chairs can allocate such a high number of instructors for legitimate reasons (cf.~Section~\ref{sec:CourseSetupLimitations}).

\subsection{Hands-On Sessions}
\label{sec:hands-on}
We designed three hands-on sessions to prepare our students for designing a small-scale empirical study.
Each session focused on a distinct programming activity that features prominently in current research on AI coding assistants: code generation, code explanation and refactoring, and debugging. 
Students worked on targeted tasks with and without GitHub Copilot, reflected on their experiences, and discussed takeaways to surface potential research questions.
We summarize each session’s setup below.

\paragraph{Session 1: Code Generation}
The first session was concerned with the correctness and quality of AI-generated code~\cite{Liu2023, Liu2024}, highlighting challenges such as the need for multi-turn prompting~\cite{Liu2024} and the risk of code smells in generated output~\cite{Siddiq2024}.
We contrasted AI-assisted solutions with manually authored code.
Students were split into two groups and given the same specification: One group implemented it with assistance from GitHub Copilot, the other without.
After $30$ minutes, students completed a short post-task questionnaire.
The groups then crossed over to a second specification for another $30$ minutes, reversing the availability of AI assistance.
Both specifications came with a test suite to verify baseline conformance.
A final questionnaire concluded the activity.
We treated this programming exercise as a small-scale experiment itself: The questionnaires provided comparative data between the two groups, which we analyzed together with the students.
The session closed with a meta-level discussion of what had been measured, what research questions such data could answer, and how such setups illustrate the structure of an empirical study.

\paragraph{Session 2: Code Explanation and Refactoring}
AI coding assistants support tasks beyond simple code generation, such as code explanation and refactoring~\cite{Nam2024,Leinonen2023}.
We again divided students into two groups and provided an already implemented project.
Their task was to understand how the project worked; one group could use an AI assistant, the other could not.
After $30$~minutes, students completed a short questionnaire.
The groups then switched conditions on a second project for another $30$~minutes, followed by another questionnaire.
As in the first session, we framed the activity as a small-scale experiment, enabling a direct comparison between AI-assisted and manual approaches.
In the debrief, we discussed the choices involved in operationalizing (e.g., productivity and code quality) and how different instruments might capture such constructs.

\paragraph{Session 3: Debugging}
The third hands-on session focused on debugging.
Students received one larger project with a test suite and had $60$ minutes to identify and fix seeded bugs.
In contrast to the previous sessions, all students were permitted to use AI assistants throughout the task.
The session concluded with a discussion on how such a setup could be studied empirically, highlighting threats to validity (e.g., defect difficulty, prior experience) and possible strategies for data collection and analysis.

\subsection{Additional Instructive Sessions}
We offered four instructional sessions:
(1) introduction to GitHub Copilot, covering student access, integration in Visual Studio Code, and available functionality;
(2) literature search strategies for finding and accessing related work;
(3) principles of empirical study design, with a focus on controlled experiments, were introduced to frame the hands-on sessions and guide students' own study planning according to best practices; and
(4) scientific presentation practices, including dos and don'ts for effective scientific talks.
These sessions were designed as concise complements to existing, more comprehensive university courses (e.g., on empirical study design), ensuring that all students, regardless of prior knowledge, were sufficiently prepared to complete the seminar.

\subsection{Empirical Studies}
\label{sec:CourseEmpiricalStudies}
Each student designed a small-scale empirical study lasting $25$--$30$~minutes.
Grading requirements at our university necessitated the decision that each student develop their own study design independently rather than in a team.
Students first submitted a draft study design, then conducted a pilot with a peer while reciprocally participating as a subject.
They had two weeks to revise their design and submitted a final version accompanied by a brief rationale that documented changes and lessons learned from the first pilot.
A second pilot followed, allowing further refinement before a hypothetical final study conduct (not part of the seminar).

To provide focus and comparability, we offered five topic areas.
Students indicated preferences, and we assigned topics accordingly.
The areas reflected different dimensions of AI coding assistants at the time: (1) \emph{code explanation and comprehension}, (2) \emph{prompt engineering and interaction with coding assistants}, (3) \emph{correctness and quality of AI-generated code}, (4) \emph{the educational impact of coding assistants}, and (5) \emph{AI in practice: professional use and workflow integration}.
For each topic, we provided a reading list to help students ground their research questions in existing literature.

\subsection{Final Presentations}
\label{sec:presentations}
Students prepared a $12$-minute presentation in which they introduced their study design, explained how they refined it across the two pilot iterations, and shared preliminary insights from exploratory analyses of the pilot data.
Importantly, they were expected to critically reflect on both the potential and the limitations of their data---acknowledging, for example, that results based on two participants cannot substantiate strong quantitative claims but can nevertheless reveal design flaws or suggest promising directions.
Furthermore, we asked students to reflect on their experience of working with an empirical approach.

Each talk was followed by a substantive discussion with peers and instructors.
These discussions clarified design details and probed the reasoning behind methodological choices, the adequacy of measurement instruments, and the interpretation of pilot findings.

\subsection{Seminar Paper}
Finally, students submitted a $15$-page, single-column seminar paper describing their study.
The paper followed a standard research-paper structure (i.e., introduction, background, related work, study design, and conclusion) and additionally documented design learnings, changes after the first pilot, and initial results from pilot participants, with brief implications for a future full study.
Beyond being a required deliverable, the seminar paper aimed to give students hands-on practice in scientific writing and proper citation, complementing the oral presentation by allowing them to exercise precise and accurate scientific communication.
For institutions with shorter seminar durations, this written report could be treated as an optional component, while still achieving the core learning objectives of the seminar, particularly the focus on empirical thinking.

\section{Teacher's Perspective: Classroom Impressions}
\label{sec:impressions}

In this section, we report our impressions on teaching this seminar, in particular, how students interacted with the material and how they conducted their studies.

\subsection{Strong Interest in the Course}

\begin{figure}
    \centering
    \includegraphics[width=1\linewidth]{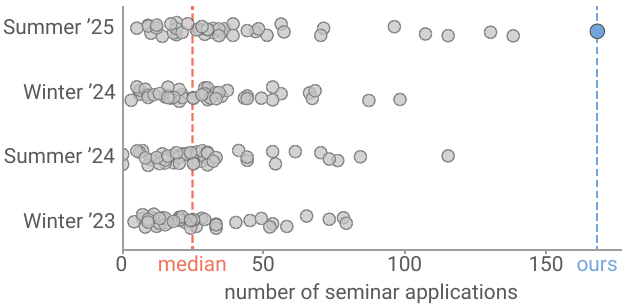}
    \caption{Number of student applications for each seminar offered over the past four semesters, highlighting our seminar with 168 applications.}
    \Description{Plot showing the number of student applications per seminar across four semesters (summer 2025, winter 2024, summer 2024, winter 2023). Each dot represents one seminar, grouped by semester on the y-axis and number of applications on the x-axis. One dot highlights the described seminar with 168 applications. A reference line indicates the median of 25 applications.}
    \label{fig:SeminarApplications}
\end{figure}

Since the computer science faculty at Saarland University employs a centrally organized seminar system, it is very transparent what seminars are offered and to gauge student interest in them. We offered 18~spots in our seminar, but demand far exceeded our capacity, with a total of 168~student applications.

To put this into perspective, we compare this number with data for the last four semesters at our faculty. Across the last four semesters, around 45 seminars were offered each semester for a total of 176 seminars. Each received a decent number of applications (median: 25, mean: 33). Our seminar far exceeded the average level of interest, with 168 applications, the highest number of recorded applications across these four semesters. This underlines how the excitement surrounding AI coding assistants is intriguing to students. We visualize all application numbers in~\autoref{fig:SeminarApplications}.

Our description of the seminar clearly communicated that the course will not just cover AI coding assistants, but to approach them with an empirical mindset. We have offered seminars covering empirical methods in the past; however, with less measurable interest. The seminar applications, which include students' motivation for participation, further highlight that students focused on the hype topic, in particular the recency of the AI assistant topic and its persuasive effects. For example, two students included the following phrases in their motivation:

\begin{itemize}[leftmargin=1em]
    \item \textit{\enquote{This seminar is very important to me as it will help me to stay up to date with the latest advancements in my research area, receive guidance from lecturers, and encounter other students' opinions.}}
    \item \textit{\enquote{I see that AI is becoming a more and more important topic in software engineering, and I find it interesting how its integration influences my personal coding experience and whether it provides an advantage in practice.}}
\end{itemize}

Overall, we not only met the expected perceived hype around the topic for our students, but it far exceeded our expectations.

\Finding{The seminar's popularity was seemingly driven by the timely and relevant topic of AI coding assistants, demonstrating that students are interested in exploring hype topics in an academic setting, even when they involve complex and abstract concepts, such as empirical methods.}

\subsection{Reflection during Hands-On Sessions}

\begin{table*}[t]
\caption{Overview of a small subset of gathered quantitative reflection data during the first hands-on session on code generation. The session included two tasks. For both tasks, we split the students into a without-AI and with-AI group.}
\label{tab:HandsOnReflection}

\centering
\fontsize{7pt}{10pt}\selectfont

\begin{tabular}{@{}l|cccc@{}}
\toprule
\multicolumn{1}{c|}{Question} & \multicolumn{2}{c}{Task 1} & \multicolumn{2}{c}{Task 2} \\
& Without AI & With AI & Without AI & With AI \\
\midrule
Q\textsubscript{R-1}: Task Completion Rate & 71\% & 50\% & 13\% & 60\% \\
Q\textsubscript{R-2}: Do you think being allowed to use GitHub Copilot (would have) made this task easier? & 100\% Yes & 100\% Yes & 100\% Yes & 100\% Yes \\
Q\textsubscript{R-3}: Did you rely on GitHub Copilot for code completion? (strongly disagree \textcolor{ColorStronglyDisagree}{\rule{1ex}{1ex}} 
\textcolor{ColorSomewhatDisagree}{\rule{1ex}{1ex}}
\textcolor{ColorNeutral}{\rule{1ex}{1ex}}
\textcolor{ColorSomewhatAgree}{\rule{1ex}{1ex}}
\textcolor{ColorStronglyAgree}{\rule{1ex}{1ex}} strongly agree) & --- &\includegraphics[width=1.8cm]{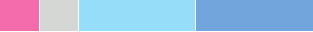} & --- & \includegraphics[width=1.8cm]{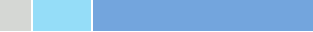} \\
Q\textsubscript{R-4}: How often did GitHub Copilot provide accurate code completions? (never \textcolor{ColorStronglyDisagree}{\rule{1ex}{1ex}} 
\textcolor{ColorSomewhatDisagree}{\rule{1ex}{1ex}}
\textcolor{ColorNeutral}{\rule{1ex}{1ex}}
\textcolor{ColorSomewhatAgree}{\rule{1ex}{1ex}}
\textcolor{ColorStronglyAgree}{\rule{1ex}{1ex}} almost always) & --- & \includegraphics[width=1.8cm]{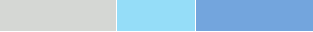} & --- & \includegraphics[width=1.8cm]{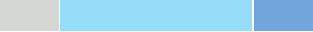} \\
Q\textsubscript{R-5}: Were there any instances where GitHub Copilot provided unexpected code completions? & --- & 38\% Yes & --- & 50\% Yes \\
Q\textsubscript{R-6}: Were there any instances where GitHub Copilot provided incorrect code completions? & --- & 63\% Yes & --- & 60\% Yes \\
\bottomrule

\end{tabular}
\end{table*}

At the beginning of the seminar, we conducted three hands-on sessions (cf.~Section~\ref{sec:hands-on}). For each of the hands-on sessions, we aimed for students not only to obtain practical experience with AI coding assistants in a controlled environment, but also to actively reflect on their experience. To encourage students to reflect on several important aspects of AI coding assistants (e.g., helpfulness, productivity, accuracy)---including tasks where the AI coding assistants were unavailable---we designed a small questionnaire with a mix of quantitative and qualitative questions.
After the first session, we asked all students to fill out this questionnaire. While the students were working on the second task, one instructor analyzed the questionnaire data. We then summarized the results as a basis for the group discussion at the end of the hands-on session.
We provide a subset of the quantitative data in~\autoref{tab:HandsOnReflection}.

Overall, the results of the questionnaire for the first session (code generation) were mixed across the two tasks. Notably, for the first task, the group without AI assistance completed more subtasks than the ones with AI (Q\textsubscript{R-1}). Nevertheless, all students across both groups reported that they would be more productive with AI coding assistants (Q\textsubscript{R-2}). While students also relied on GitHub Copilot for code completions (Q\textsubscript{R-3}) and perceived it as fairly reliable (Q\textsubscript{R-4}), they also reported that it often made unexpected (38--50\%, Q\textsubscript{R-5}) or incorrect (60--63\%, Q\textsubscript{R-6}) code completion suggestions. These contradicting experiences, which also appear to be quite task-specific, were part of the in-class discussions. The open-text questions about the students' experience similarly covered negative and positive experiences, such as:
\begin{itemize}[leftmargin=1em]
    \item \textit{\enquote{I tried commenting while Copilot was editing... bad idea, the edit didn't make any sense.}}
    \item \textit{\enquote{Copilot did basically everything [...] it was completely helpful}}
\end{itemize}

For the second task in the first session, the group using AI coding assistants achieved a notably higher completion rate, suggesting higher productivity. Otherwise, results were similar regarding perceived usefulness of the assistants and their incorrect completions.

We provided students with a test suite as an indicator of task completion but did not disclose additional secret tests covering edge cases. We intuitively expected that students coding without AI assistants would develop a deeper understanding and thus miss fewer edge cases than those relying on generated code. However, because students progressed at different rates within the limited time, the hidden test results were not comparable between groups.

For the second and third sessions, students could still use the same questionnaire for personal reflection, but we did not demand it in a structured manner on a classroom level. Instead, we employed a live group discussion covering the same topics without presenting specific data, since students were already familiar with the aspects that we considered worthy of a discussion.

\Finding{During the hands-on sessions, we encouraged students to reflect on their experiences with AI coding assistants via a questionnaire, which provided insights into their perceptions and experiences. It also facilitated a group discussion, in which students could share their thoughts and observations about using these tools in our scenario.}

\subsection{Conducting Empirical Studies}

After the hands-on sessions, it was the students' turn to design their own empirical study on one small aspect of AI coding assistants.

\subsubsection{Students' Research Goals and Methods}

We assigned the students to five different research areas~(cf.~Section~\ref{sec:CourseEmpiricalStudies}). This way, we enforced variety in the students' topics. However, while we did not restrict students regarding their choice of research method, the chosen methodology was unexpectedly very uniform across our students. Except for one student, all students opted for some kind of (controlled) experiment. This was not our intention, as we did not restrict the methods and let students freely choose. We suspect that several factors may drive this trend and make other methods less feasible in our format. Specifically, our hands-on sessions were already in a style of controlled experiment. Furthermore, our student pilot study setup assigned each student one participant for half an hour. We paired the students together, which means they could not search for participants with a specific profile themselves. For future seminars, in which different methods would be essential, a different pilot study setup should be considered.

In combination with a controlled experiment, many students used a post-questionnaire to collect self-reported perspectives alongside behavioral measures. Again, this approach may have been strongly inspired by our hands-on sessions, in which we also combined a practical task with reflection driven by a questionnaire, and explored differences in behavior and subjective viewpoints. Furthermore, many students employed comparative setups (e.g., prompting style A vs. prompting style B, or with and without AI coding assistants), again similar to our hands-on sessions.

While the methods were fairly uniform, the conceptual goals differed between students and fully covered our intended variety of research topics. The students explored different research topics within the assigned areas, 
from very technical topics (e.g., ``Which prompting style produces more correct code?'' or ``To what extent does GitHub Copilot generate secure and correct API implementations using FastAPI and PostgreSQL?'') to their practical effects (e.g., ``How reliable is code generated by GitHub Copilot perceived by developers?''), and to their role in a learning environment (e.g., ``Do AI-coding tools help novice programmers learn a new language?'').
From our perspective, this variety is quite valuable, as students can learn not just from their own study but also benefit from the insights of their fellow students.
Retrospectively, we have to note that the seminar structure does make some topics less feasible. For example, generally interesting research topics, such as using AI coding assistants in a collaborative setting or the effects of AI coding assistants over an extended period of time, were not feasible. If this is of interest, it might be worth considering refinements for future iterations of this seminar~(cf.~Section~\ref{sec:CourseSetupLimitations}).

\paragraph{Evolution of Study Designs}

Students first created an initial study design, conducted a pilot, and then refined their design before a second pilot. 
We reviewed the students' submissions and analyzed the changes between their initial experiment design and the refined version. While we expressed that students only needed to refine the study design, many also improved their writing clarity and presentation quality. In retrospect, this made it difficult to separate presentation gains from methodological improvements. To more clearly capture this learning step, future iterations should require an explicit section detailing (1) what changed and (2) why.

Across all students, we observed a wide range of revisions. While a couple of students only applied minor tweaks, many substantially improved their studies---e.g., reducing task complexity, or introducing randomization. Notably, a few students overhauled their designs, changing the population (from targeting experienced programmers to novice programmers), experimental setup (different programming language), or even the research question. Overall, we were impressed with students quickly being able to refine their study designs after the first pilot, despite little prior experience. This aligns with prior findings that small projects~\cite{Felderer2019} and practical hands-on work are often relevant for teaching empirical software engineering~\cite{Luz2022}. In our case, all students demonstrated that they can pick up one key aspect of empirical research---adapting and improving a methodology in response to unexpected challenges or practical limitations---deepening their understanding of the complexities involved in designing and conducting empirical studies.
Our live observations of students conducting pilot studies confirmed our analysis of the study design documents. 

\paragraph{Reflection of Learnings from Conducting Studies}

In their final presentations, students shared both topic-related insights (e.g., ``LLM explanations give answers, not insights'') and reflections on their empirical work.
Many discussed strengths and weaknesses of their designs (e.g., ``Within-subject counterbalancing minimized individual and learning biases''), challenges (e.g., ``measuring \emph{understanding} was hard''), and ideas for improvement (e.g., ``prepare a scripted set of interview questions in advance'').
Some also emphasized the importance of considering confounding factors early on.
These are all valuable considerations even for experienced researchers.
 
\Finding{Our seminar allowed students to design and conduct their own small-scale empirical studies. While the setup may have limited the range of methods, the two-pilot design proved valuable by supporting refinement and improving study quality. Many students also reported gaining key insights into the challenges of empirical research.}

\section{Student Perspective: Linking Hype to Empirical Thinking}
\label{sec:survey}

To complement our subjective impressions, we aimed to understand how students perceived the seminar and how engaging with a currently popular topic influenced their learning experience. Thus, we invited all students to participate in a voluntary online survey following the final presentations.
Note that while our primary goal is to provide an experience report, we complement it with small-scale empirical evaluations to strengthen the insights gained.

\subsection{Methodology}

To ensure scientific rigor, we follow \citeauthor{Kasunic:2005:Survey}’s seven-step guideline for survey design~\cite{Kasunic:2005:Survey}, complemented by \citeauthor{Linaker:2015:Survey}'s more recent annotations~\cite{Linaker:2015:Survey}.
We collect both qualitative data, which provide detailed insights into individual participants’ experiences~\cite{Melegati:2024:QualitativeSurveys}, as well as quantitative data, which---despite the small sample size---are analyzed exploratively to offer a more general indication of the seminar's effectiveness.

\paragraph{Sampling Strategy} 
After the final presentations (cf.~Section~\ref{sec:presentations}), we shared the survey link with all seminar participants and invited them to take part without offering incentives.
They could either use the remaining time of the seminar session to complete the survey on site, fill it out at home, or choose not to participate at all.
The survey was anonymous, and participation was entirely voluntary.
In line with standard practice at Saarland University, this anonymous course evaluation did not require institutional review board approval.

\paragraph{Questionnaire Design} 
The questionnaire began with a consent form that informed participants about the purpose of the survey and the data processing procedures.
The questionnaire consisted of four main sections spread across four pages.
The first section focused on students' perceptions of AI coding assistants, prompting them to reflect on their views before and after the seminar.
The second section assessed the achievement of empirical thinking goals by asking students whether and how they adopted an empirical mindset throughout the course.
The third section explored the link between working on a currently popular topic and students' learning outcomes, for example, by asking how engaging with a trending technology such as AI coding assistants influenced their interest in designing and conducting their own study.
The final section gathered demographic information, including participants' study program, current semester, prior experience with AI coding assistants, and prior experience with conducting empirical research.

\paragraph{Pilot Test and Data Collection} 
It is generally recommended to pilot surveys~\cite{Kasunic:2005:Survey}. However, since our intended sample coincides entirely with the population of students who can meaningfully comment on the seminar, we opted against a traditional pilot. Instead, we asked two doctoral students familiar with the seminar to review the questionnaire and included their feedback in the final version. 
The survey was online for 14 days.
In total, 11 of 13 students started the survey, of which 10 students completed the survey.

\paragraph{Demographic Data}

\begin{table}[]
\caption{Overview of our student demographics and experience with AI coding assistants and empirical research.}
\label{tab:demographics}

\fontsize{8pt}{10pt}\selectfont
\begin{tabular}{llr}
\toprule
Measure & \multicolumn{2}{r}{No./Mean $\pm$ SD} \\
\midrule
\multicolumn{2}{l}{Number of students completing the seminar} & 13 \\
\multicolumn{2}{l}{Number of students who completed the survey} & 10 \\
\specialrule{0.1pt}{1pt}{1pt}
Undergraduate students & & 4 \\
$\rightarrow$ Current semester & & 5.25 $\pm$ 0.83 \\
Graduate students & & 6 \\
$\rightarrow$ Current semester & & 1.66 $\pm$ 0.75 \\
\specialrule{0.1pt}{1pt}{1pt}
  & None & 2 \\
Experience with & Tried once or twice & 1 \\
AI coding assistants & Used occasionally & 3 \\
& Used regularly & 4 \\
\specialrule{0.1pt}{1pt}{1pt}
 & None & 4 \\
Experience with & Minimal & 4 \\
Empirical Research & Some & 2 \\
& Solid & 0 \\
\bottomrule
\end{tabular}
\end{table}

The survey included typical demographic and experience questions, which we present in~\autoref{tab:demographics}.

\paragraph{Quantitative Data Analysis}

To analyze the quantitative data from the Likert-scale~\cite{Likert1932} questions of the survey, we mapped the response data to the respective categories (e.g., ``strongly agree'', ``disagree''). We then computed descriptive statistics in terms of mean and standard deviation (SD). Note that we are aware of an open debate on whether Likert scales represent ordinal or continuous intervals~\cite{murray2013likert} and that we opted to consider Likert items as discrete values on a continuous scale. We provide the results, including visualizations, in~\autoref{tab:SeminarReflectionQuantitative}.

\paragraph{Qualitative Data Analysis}

To analyze the qualitative data from the open-ended survey questions, we conducted a lightweight coding and thematic analysis, roughly following the recommendations and steps of Cruzes and Dyba~\cite{Cruzes2011} and Defranco and Laplante~\cite{Defranco2017}. Specifically, we conducted the analysis in the following steps:

\begin{enumerate}[leftmargin=2em]
    \item \textbf{Prepare data}: We extracted all answers for each question.
    \item \textbf{Code data}: One author initially coded each available answer. This was done with an open, inductive coding approach due to the novelty of the questions and the small dataset.
    \item \textbf{Create themes from codes}: The same author then found common themes without a hierarchy across codes.
    \item \textbf{Review coding and themes}: Another author reviewed and checked the coding, themes, and interpretation. They also assisted in discussing difficult cases.
    \item \textbf{Assess overall quality}: A third author conducted a final quality assessment based on Treude's quality criteria~\cite{Treude2024}.
\end{enumerate}

We discuss the results and our interpretation in the next section. For full transparency, we provide all answers with assigned codes and themes in our replication package.

\subsection{Results}

We structured the content-related part of the survey into three sections, each covering a different topic. We discuss each section in the following.

\begin{table*}[th!]

\setlength{\fboxsep}{1pt}
\caption{Overview of gathered quantitative reflection data of the post-seminar survey. All questions are on a 1$\rightarrow$5 Likert-scale~\cite{Likert1932} from \textcolor{ColorStronglyDisagree}{\rule{1ex}{1ex}} ``strongly disagree'' to \textcolor{ColorStronglyAgree}{\rule{1ex}{1ex}} ``strongly agree''. We excluded ``I do not know'' answers.}
\label{tab:SeminarReflectionQuantitative}

\centering
\fontsize{7pt}{10pt}\selectfont

\begin{tabular}{@{}ll|ccr@{}}
\toprule
& \multicolumn{1}{c}{Question} & Mean $\pm$ SD & Response Distribution \\
\midrule
\multirow{8}{*}{\rotated{(a) AI coding assistants}} & Q\textsubscript{AI-L1}: Before the seminar, I had a clear idea of what AI coding assistants can and cannot do &  3.50 $\pm$ 1.27 & \TablePlot{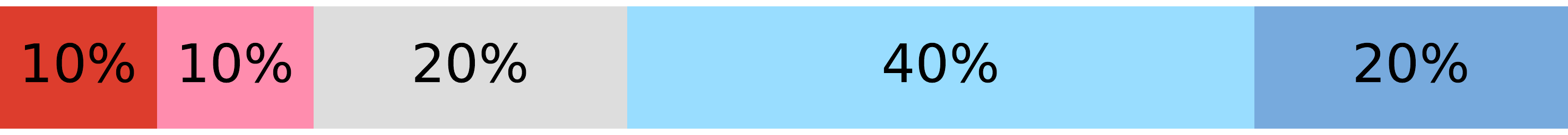} \\
& Q\textsubscript{AI-L2}: Before the seminar, I perceived AI coding assistants as accurate and reliable tools & 3.20 $\pm$ 0.92 & \TablePlot{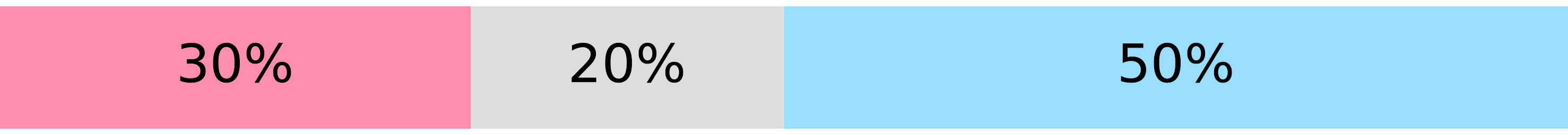} \\
& Q\textsubscript{AI-L3}: After the seminar, I perceived AI coding assistants as accurate and reliable tools &  3.50 $\pm$ 0.97 & \TablePlot{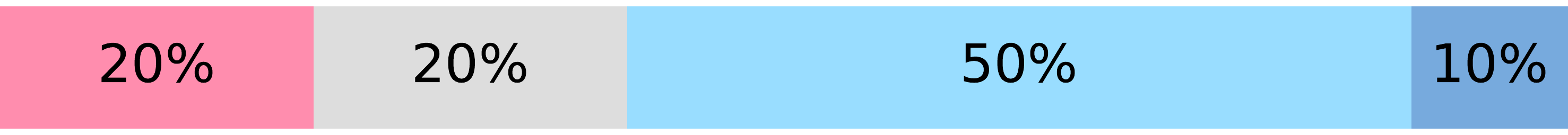} \\
& Q\textsubscript{AI-L4}: During this seminar, my understanding of the capabilities and limitations of AI coding assistants changed & 4.00 $\pm$ 1.05 & \TablePlot{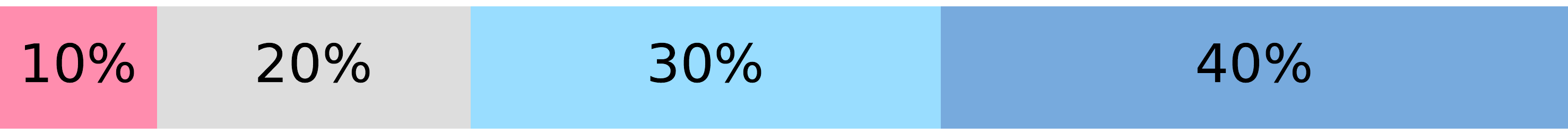} \\
& Q\textsubscript{AI-L5}: After this seminar, I better understand both the strengths and limitations of AI coding assistants & 4.50 $\pm$ 0.71 & \TablePlot{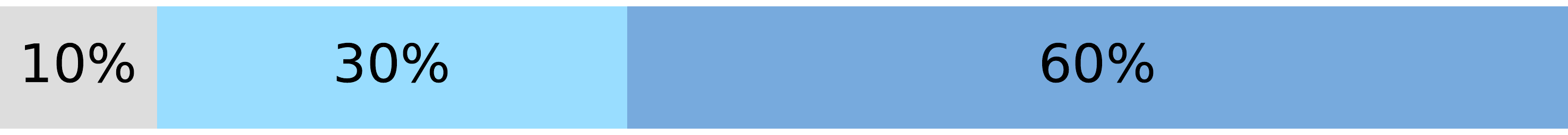} \\
& Q\textsubscript{AI-L6}: After the seminar, I am more likely to critically evaluate the output of AI coding assistants & 4.20 $\pm$ 0.63 & \TablePlot{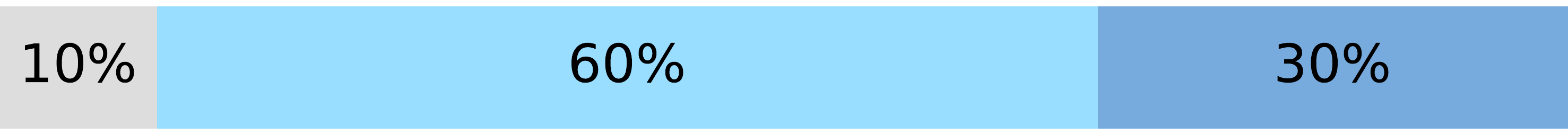} \\

\specialrule{0pt}{8pt}{1pt}

\multirow{12}{*}{\rotated{(b) Empirical Mindset}} & Q\textsubscript{Emp-L10}: Helped to distinguish between anecdotal and empirical evidence & 4.00 $\pm$ 0.71 & \TablePlot{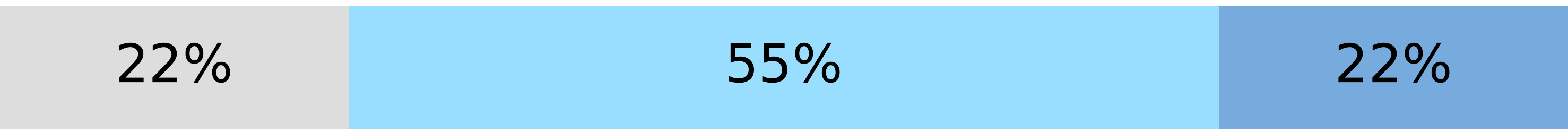} \\
& Q\textsubscript{Emp-L11}: More likely to seek out comprehensive data and robust evidence & 4.20 $\pm$ 0.63 & \TablePlot{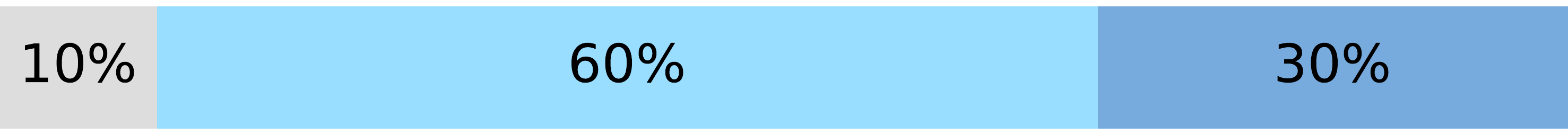} \\
& Q\textsubscript{Emp-L12}: More confident in assessing claims about the capabilities of AI coding assistants & 4.20 $\pm$ 0.92 & \TablePlot{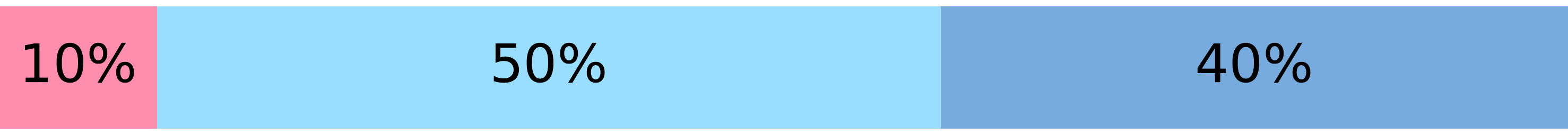} \\
& Q\textsubscript{Emp-L13}: More confident applying an empirical mindset to other emerging technologies & 4.20 $\pm$ 0.63 & \TablePlot{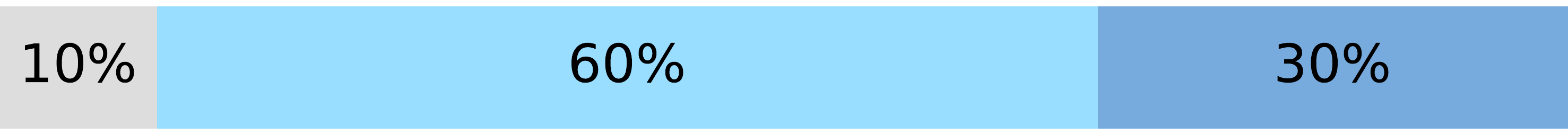} \\

& Q\textsubscript{Emp-L14}: My own empirical study helped me better understand how empirical research works & 4.70 $\pm$ 0.48 & \TablePlot{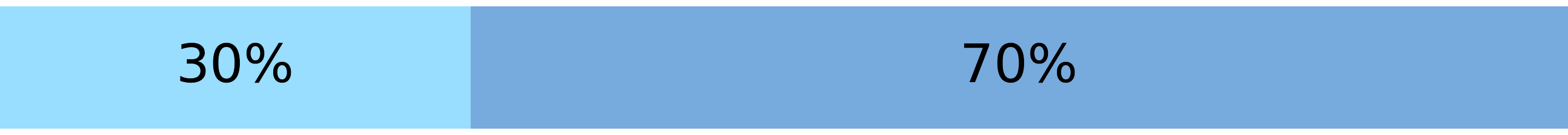} \\
& Q\textsubscript{Emp-L15}: My own study made me feel invested in the results and motivated to share my findings & 4.50 $\pm$ 0.53 & \TablePlot{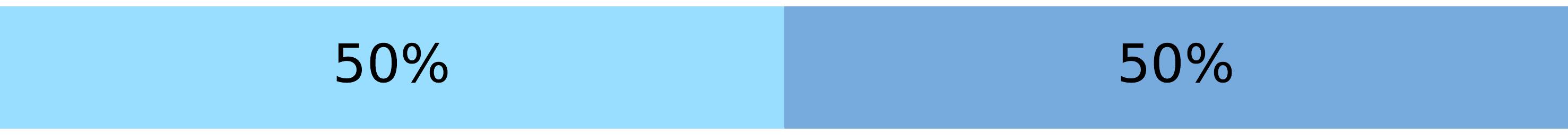} \\
& Q\textsubscript{Emp-L16}: Individual supervision played an important role in guiding my empirical work & 4.00 $\pm$ 0.94 & \TablePlot{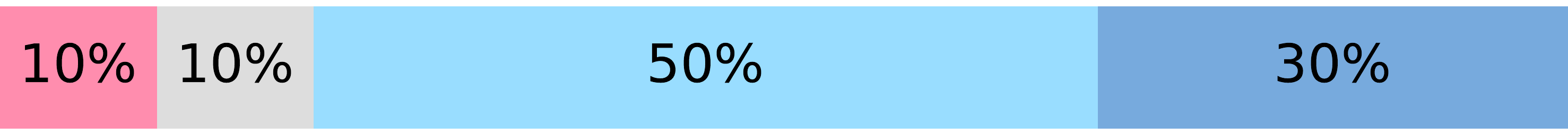} \\
& Q\textsubscript{Emp-L17}: Questions from my peers helped me reflect on methodological issues with my study design & 4.10 $\pm$ 0.74 & \TablePlot{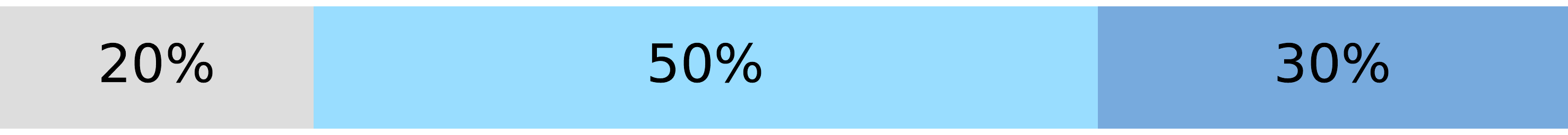} \\
& Q\textsubscript{Emp-L18}: Discussing my study improved understanding more than theory alone & 4.20 $\pm$ 0.79 & \TablePlot{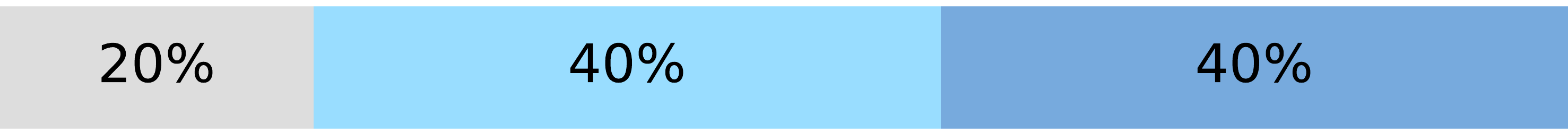} \\

\specialrule{0pt}{8pt}{1pt}

\multirow{5}{*}{\rotated{(c) Linking Hype}} & Q\textsubscript{Hype-L21}: Conducting a study on a hyped topic motivates me more than working on a conventional topic & 3.80 $\pm$ 0.79 & \TablePlot{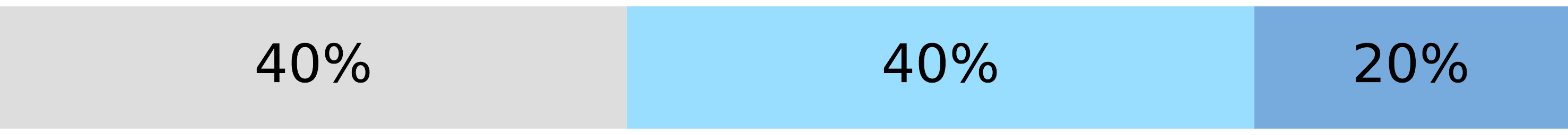} \\
& Q\textsubscript{Hype-L22}: Working on AI coding assistants helped me better understand the value of empirical investigation &  4.10 $\pm$ 0.74 & \TablePlot{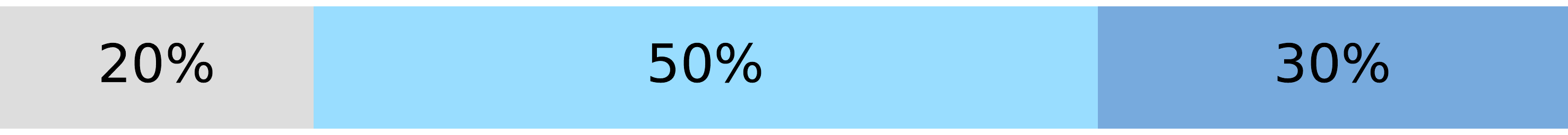}\\
& Q\textsubscript{Hype-L23}: A predefined topic area with provided literature eased developing a feasible study design & 4.30 $\pm$ 0.67 & \TablePlot{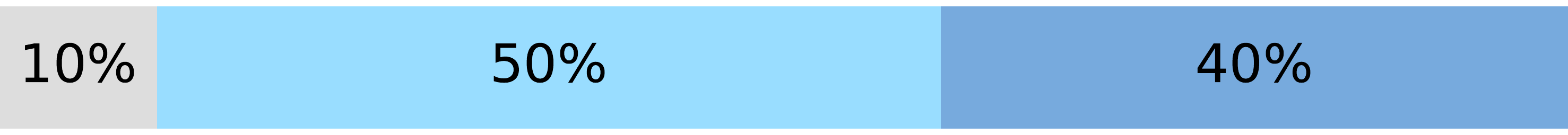} \\
& Q\textsubscript{Hype-L24}: Conducting own research deepened my critical understanding more than traditional lectures & 4.30 $\pm$ 0.82 & \TablePlot{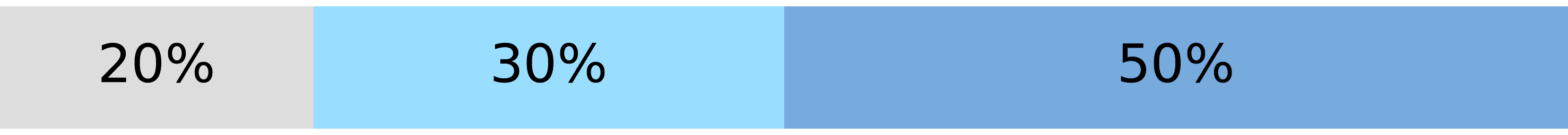} \\

\bottomrule
\end{tabular}
\end{table*}

\subsubsection{Perception of AI Coding Assistants}

\begin{table}[h!]
\caption{Themes from answers to open-ended questions in three topic areas of our post-seminar survey. Frequency denotes the count of participants contributing to a given theme.}
\label{tab:SeminarReflectionQualitative}

\fontsize{8pt}{10pt}\selectfont
\begin{tabular}{lr}
\toprule
\textbf{(a) Themes on AI Coding Assistants} & \textbf{Frequency} \\
\specialmidruleTop

\multicolumn{2}{c}{\cellcolor{ColorGrayBackground}\begin{tabular}[c]{@{}c@{}}\textit{Q\textsubscript{AI-O7}: How has your perception of AI coding assistants changed}\\ \textit{throughout the seminar?}\end{tabular}} \\
\specialmidruleBottom
Limitation and Risk Awareness &  6/8 \\
Usefulness When Used Carefully &  6/8 \\
Realizing Drawbacks for Student Learning &  2/8 \\
Awareness of Research Gaps &  1/8 \\

\specialmidruleTop
\multicolumn{2}{c}{\cellcolor{ColorGrayBackground}\begin{tabular}[c]{@{}c@{}}\textit{Q\textsubscript{AI-O8}: Were there particular hands-on session(s) that changed}\\ \textit{how you see AI coding assistants?}\end{tabular}} \\
\specialmidruleBottom
Confirming/Positive Response &  6/8 \\
Discovering Features and Use Cases &  6/8 \\
Experience Limitations & 3/8 \\
Negative Response &  2/8 \\

\specialmidruleTop
\multicolumn{2}{c}{\cellcolor{ColorGrayBackground}\begin{tabular}[c]{@{}c@{}}\textit{Q\textsubscript{AI-O9}: Was there a particular part of your own study that}\\ \textit{changed how you see AI coding assistants?}\end{tabular}} \\
\specialmidruleBottom
Change in Perception (e.g., Quality) & 5/8 \\
Human Factor in Tool Effectiveness &  2/8 \\
Influence of Prompt Engineering &  2/8 \\
No Change &  2/8 \\

\specialrule{0pt}{8pt}{1pt}
\textbf{(b) Themes on Empirical Thinking} & \textbf{Frequency} \\
\specialmidruleTop

\multicolumn{2}{c}{\cellcolor{ColorGrayBackground}\begin{tabular}[c]{@{}c@{}}\textit{Q\textsubscript{Emp-O19}: What challenges did you encounter when trying }\\ \textit{to apply an empirical approach to studying AI coding assistants?}\end{tabular}} \\
\specialmidruleBottom
General Empirical Difficulties & 4/8 \\
Seminar-Specific Difficulties & 4/8 \\
Technical Issues & 1/8 \\

\specialmidruleTop
\multicolumn{2}{c}{\cellcolor{ColorGrayBackground}\begin{tabular}[c]{@{}c@{}}\textit{Q\textsubscript{Emp-O20}: How did conducting a first pilot study }\\ \textit{influence your expectations or plans for the final study design?}\end{tabular}} \\
\specialmidruleBottom
Challenges in Running Experiments & 7/8 \\
Experiment Redesign/Reduction & 5/8 \\
Human Variability in Software Studies & 4/8 \\
Reducing Uncertainty & 3/8 \\

\specialrule{0pt}{8pt}{1pt}
\textbf{(c) Themes on Linking Hype} & \textbf{Frequency}\\
\specialmidruleTop

\multicolumn{2}{c}{\cellcolor{ColorGrayBackground}\begin{tabular}[c]{@{}c@{}}\textit{Q\textsubscript{Hype-O25}: How did working on a current topic of AI coding assist-}\\ \textit{ants affect your interest in designing and conducting your own study?}\end{tabular}} \\
\specialmidruleBottom

Hype Increased Interest & 5/8 \\
Personal Relevance and Learning & 2/8 \\
Empirical Learning/Research  & 2/8 \\

\bottomrule
\end{tabular}
\end{table}

Our first section of the survey investigated the students' diverse prior knowledge (Q\textsubscript{AI-L1}) and perception (Q\textsubscript{AI-L2}) of AI coding assistants, and how it changed throughout this seminar (cf.~\autoref{tab:SeminarReflectionQuantitative}). Initially, students were hesitant about the accuracy and reliability of AI coding assistants (Q\textsubscript{AI-L2}), which shifted slightly more positive throughout the seminar (Q\textsubscript{AI-L3}). The majority of our students reported that our seminar has helped them to understand the capabilities and limitations of AI coding assistants (Q\textsubscript{AI-L4}, Q\textsubscript{AI-L5}) and that they are more likely to critically evaluate them (Q\textsubscript{AI-L6}).

Analyzing the open-ended questions supported this viewpoint, which we show in~\autoref{tab:SeminarReflectionQualitative}(a). Throughout the seminar, the majority of students report that they have developed a more comprehensive understanding of AI coding assistants (Q\textsubscript{AI-O7}). In particular, they are more aware of their risks and limitations (S2, S4, S6, S5, S7, S9, S11), but also their selective usefulness when used appropriately (S2, S4, S7, S7, S10, S11). Two students (S4, S11) highlighted that AI coding assistants can be problematic for learning students. One student (S1) also pointed out that they are now more aware that there is still much to explore in this research field. For example, S9 states ``I started to see AI more with respect to its limitations but maybe I was getting too reliant on it and this seminar was a wakeup call'', while S11 stated ``I learned to what extent AI coding assistants can be used for learning but also where the limitations are, so where students need to make progress by themselves.''

For the majority of our students (S1, S2, S4, S6, S7, S11), the hands-on sessions were perceived as instructive (Q\textsubscript{AI-O8}). They report that across all three hands-on sessions (code generation, code explanation, debugging) some or multiple sessions provided specific learnings. They also report that they discovered new features and explored use cases they had not considered before. For example, S6 states: ``I know that Copilot can directly edit our entire code file, and we can chat with the assistant. Especially helpful when asking it to explain the function by providing example input and output''. Some students (S4, S7, S11) experienced firsthand the limitations of AI coding assistants, such as incorrect suggestions. Only a couple of students (S9, S10) stated that the sessions had no substantial impact on how they perceived AI coding assistants. 

Through their own studies, many students (S1, S2, S7, S9, S10) reported an increased awareness of AI coding assistants’ strengths and limitations (Q\textsubscript{AI-O9}). In particular, they discussed their fast development, discovered new use cases, but also noted potential issues. Some students (S4, S10) emphasized the impact of the human factor, in particular that other students use AI coding assistants differently, and this strongly influences their success and perception. 
Two students (S2, S7) discussed that prompt engineering is a key factor to successfully use AI coding assistants. Two different students (S6, S11) reported no substantial change in perception toward AI coding assistants from conducting their own studies.

\subsubsection{Empirical Thinking}

While the first section of our survey confirmed the learning effects regarding AI coding assistants, one fundamental goal was to teach empirical thinking. We devised 9 questions targeting various aspects of how participating in our seminar affected students' empirical thinking in general and specifically regarding AI coding assistants. The overwhelming majority consistently agreed or strongly agreed with the positive effects of our seminar (Q\textsubscript{Emp-L7}--Q\textsubscript{Emp-L15}, cf.~\autoref{tab:SeminarReflectionQuantitative}(b)).

For our open-ended questions (cf.~\autoref{tab:SeminarReflectionQualitative}(b)), students encountered various difficulties when conducting their empirical pilot studies (Q\textsubscript{Emp-O19}). Half of the responses (S2, S5, S6, S7) covered typical challenges of empirical research, such as defining research questions, selecting measures, and trade-offs in experiment design and validity considerations. Another set of challenges was reported due to the design and conduct of our seminar (S5, S6, S9, S10), which limited the students' research to two short pilot studies (S9: ``There were only two participants''), but also due to issues in providing sufficient information. One student (S11) reported technical difficulties but was able to overcome them.

Within the scope of our seminar, students reported a lot of refinements in their experiments (Q\textsubscript{Emp-O20}). For the majority (S1, S2, S4, S5, S7), this meant a reduction in scope by simplifying the task or reducing the number of tasks. This was partially driven by the artificial time limit we had set to stay within the seminar time slot. Students also recognized general challenges of designing empirical studies, either due to biases or due to the difficulty of the human factor in software engineering. Finally, a few students (S4, S6, S11) noted that they were unsure what to expect, and conducting the pilot studies was important for refinement.

\subsubsection{Linking Hype and Empirical Thinking}

In the final section of the questionnaire, we explored how a hype topic could be used to engage students in empirical work. Here, we received a few neutral responses, but still a majority of students agreed or strongly agreed that AI coding assistants were able to motivate them for empirical work (Q\textsubscript{Hype-L21}, Q\textsubscript{Hype-L22}). Additionally, they agreed that conducting their own study is more effective than theoretical lectures (Q\textsubscript{Hype-L24}). We initially were concerned that students would struggle with finding relevant literature for a novel research topic. Thus, we pre-defined five research areas and, for each, provided them with a list of literature as a starting point. Almost all students confirmed this was helpful for developing their studies (Q\textsubscript{Hype-L23}).

The answers to the open-text questions~(cf.~\autoref{tab:SeminarReflectionQualitative}(c)) complemented and reinforced the trends found in the Likert-scale responses. Many students (S4, S6, S7, S8, S9, S11) reported that working on the hype topic of AI coding assistants increased their interest in empirical research, especially due to its novelty and personal relevance. In particular, students mentioned: ``It made it way more interesting!'' (S4) and ``I'm not really interested in research side but conducting this study made me curious'' (S9). A few students (S2, S5, S7, S10) explicitly stated that conducting a pilot study helped them to better understand empirical research (S1: ``It taught me basically how to conduct studies''). Some students (S2, S5, S6, S9, S11) also emphasized that the research results were meaningful to them personally. Interestingly, one student (S10) stated that their interest in empirical methods was genuine and not because of AI.

\subsubsection{Final Notes}

In the final section, a few students (S2, S9, S10, S11) offered closing remarks, all highlighting positive aspects of the seminar, such as an overall enjoyable experience (S9: ``I had fun''), the seminar's structure, and the opportunity to receive feedback. Two students (S9, S10) also noted a negative aspect, namely unclear milestone expectations.

\Finding{In a nutshell, survey results, with Likert-scale responses from 10 students and open-ended feedback from 8 (of 13 total), indicate the seminar was perceived very positively. Students gained a deeper understanding of AI coding assistants and further developed their empirical thinking. Additionally, our data suggest that leveraging hype topics can effectively engage students in empirical research, especially when exploring a relevant real-world application through their pilot study.}

\section{Discussion and Lessons Learned}
\label{sec:lessons}

In this section, we reflect on our experiences teaching the seminar and highlight lessons to inform actionable advice for others.

\subsection{Hype Can Be a Pedagogical Catalyst}

Our seminar on AI coding assistants suggests that hype can be a powerful entry point to core methodological training. 
A record number of applicants and positive student feedback indicated that hype-driven topics can draw students into empirical thinking and research skills.
One benefit is that limited prior research on hype topics makes research feel more relevant: students reported being motivated by 
working in a novel area and by having to think more carefully about study design---which made the work more exciting.
Framing the seminar this way helped create a learning environment where students were motivated and invested in the material. 

However, instructors must ensure that hype does not overshadow careful methodological rigor. 
Because fads are temporary but empirical skills endure, we recommend using hype-driven topics primarily as an entry point, explicitly anchored to non-negotiable methodological learning goals.

\subsection{Managing Expectations Beyond the Hype} 

Building on the motivational benefits of hype, we also observed its risks.
From 168 applications, we admitted 18 students, of whom 13 completed their empirical studies.
Some dropout is expected, for example, due to university changes or personal circumstances, but others left without clear reasons. This raises an important question: Was there a mismatch between the expectations of these students (possibly focused on AI coding assistants) and the actual content of the seminar (with focus on empirical methods)? Hype may have attracted many applicants, but it may also have drawn in students mainly interested in \enquote{playing with} AI coding assistants rather than engaging in empirical research. 

A key lesson learned is that hype amplifies the cost of misaligned expectations.
While hype can make complex topics accessible, if not framed carefully, it can foster unrealistic expectations and student disengagement. Although our seminar description clearly stated its goals (i.e., empirical thinking, research design, and pilot studies), some students may have focused more on the topic than on this framing. This emphasizes the importance of transparent communication about workload, deliverables, and intended learning outcomes, in line with prior teaching experiences~\cite{Vilela2024}.
In practice, our experience suggests foregrounding methods, workload, and deliverables more prominently than the topic itself in course descriptions and early sessions.

\subsection{Ownership of Inquiry Fosters Insights}

When students design and carry out their own studies, they naturally want to share their findings---deepening their critical engagement with the topic~\cite{Vegas:2024:CourseExperimentation}. In our seminar, this sense of ownership was amplified through the iterative process of running two pilot studies, reflecting, and refining. Students could clearly see how their design evolved through their own decisions, which made them more invested in the outcome. Felderer and Kuhrmann found a similar effect when their students conducted small studies~\cite{Felderer2019}.

Presenting study designs to peers further created accountability and learning opportunities~\cite{Vilela2024}. Group discussions about threats to validity were more productive when grounded in actual student projects rather than hypothetical examples. This allowed students to learn not only from their own project but also from those encountered by their peers.
Overall, our experience suggests that student-driven inquiry is highly beneficial for developing empirical thinking, echoing prior work~\cite{Luz2022}, but instructors should plan sufficient feedback capacity and structured checkpoints to support students throughout the process.

\subsection{Balance of Dual-Purpose Seminars} 

While our primary learning goals focused on empirical thinking, the seminar also gave students extensive exposure to AI coding assistants. This showed that technical and research skills do not need to be taught in isolation---integrating them can make each side more meaningful. 
By combining hands-on exercises with empirical study design, we created a dual-purpose learning pathway: students practiced with AI coding assistants in realistic tasks while also reflecting on these experiences through the lens of research.

This approach was effective for students motivated by technology but less inclined toward pure methodological training. Integrating technical and research perspectives made both more meaningful and engaging. This shows that even within a limited time frame (compared to, e.g., completing a graduate program), students can meaningfully experience a well-scoped research process. The pilot studies provided a \enquote{scaled-down but authentic} research experience, mimicking the workflow of a full research project. While these pilot studies cannot replace full-scale empirical projects, they are an efficient and effective compromise for a seminar.

Teaching this seminar corroborated that frequent, short-cycle feedback is essential to ensure students develop meaningful and methodologically sound study designs~\cite{Vilela2024}. This was particularly crucial for those without prior empirical training. 
Without structured feedback and multiple interim deadlines, students risk drifting into pseudo-scientific territory.

\subsection{Limitations of Course Format}
\label{sec:CourseSetupLimitations}

The seminar offered valuable hands-on experience in empirical study design, but it came with some limitations.
Students mainly conducted two short pilot experiments, so longer-term or qualitative studies (e.g., field studies, interviews with practitioners) were not feasible.
Adapting the structure would be needed for different methods or extended investigations.
Moreover, given the participants' limited experience in empirical studies, students may not execute everything flawlessly unless provided with sufficient guidance.
In this regard, the high staff-to-student ratio helped provide individualized guidance but may be difficult to replicate in contexts with fewer instructors.
In such a case, supervision could shift toward in-class group feedback while still supporting student learning.

Finally, while designed for software engineering, similar approaches could work in other disciplines where trending topics or emerging technologies spark curiosity and can motivate students to explore questions critically.
In these fields, instructors could leverage current debates, innovations, or emerging tools to create a comparable learning environment.
By aligning hands-on engagement with structured reflection, students in any domain can be guided to critically evaluate claims, gather evidence, and develop robust, evidence-based insights.
Overall, our experience suggests that this format works best for short, tightly scoped empirical studies and may require careful adaptation when transferred to contexts with fewer instructors or different methodological goals.

\section{Conclusion}
\label{sec:conclusion}

Hype cycles are a recurring reality in software engineering, shaping industry practices, research agendas, and also what students expect to learn. 
Although hype in software engineering is rarely met with empirical scrutiny, its bold claims can serve as a powerful teaching opportunity to engage students in formulating and testing hypotheses about what these technologies can and cannot deliver.
In this experience report, we reflected on our seminar on AI coding assistants, which embraced this dynamic: We used the hype to spark interest and directed it toward deeper engagement with empirical research.
Through hands-on exercises, pilot studies, and iterative reflection, students learned to design, evaluate, and communicate empirical work, while instructors provided close guidance and structured feedback.

From our experience, we derived several lessons learned: hype can be a powerful pedagogical catalyst, but it can also raise false expectations if students primarily seek technical training rather than research practice; ownership of inquiry fosters ownership of insight; and dual-purpose formats that balance technical skills with empirical reflection can be effective.

We hope this report encourages others to treat the next hype not just as a teaching challenge, but as an opportunity to equip students with empirical thinking skills---skills that will remain essential, no matter what the next trend may be.

\begin{acks}
We thank all participating students for their curiosity, engagement, and valuable contributions to both the seminar and our accompanying research. 
This research is supported by ERC Advanced Grant \enquote{Brains On Code} (101052182) and \grantsponsor{dfg}{DFG}{https://www.dfg.de}
 through the Collaborative Research Center
 \grantnum[https://perspicuous-computing.science]{dfg}{TRR 248, project ID 389792660}.
\end{acks}

\bibliographystyle{ACM-Reference-Format}
\bibliography{main}

\end{document}